\documentclass[acmsmall]{acmart}

\AtBeginDocument{%
  \providecommand\BibTeX{{%
    \normalfont B\kern-0.5em{\scshape i\kern-0.25em b}\kern-0.8em\TeX}}}

\setcopyright{acmcopyright}
\acmJournal{TOIT}
\acmYear{2024} \acmVolume{X} \acmNumber{X} \acmArticle{1} \acmMonth{10} \acmPrice{XXX}\acmDOI{10.1145/370222}

\usepackage{amsmath,amsfonts}

\usepackage{graphicx}
\usepackage[caption=false]{subfig}

\usepackage[outdir=./]{epstopdf}
\usepackage{textcomp}
\usepackage{xcolor}
\usepackage{url}

\usepackage{enumitem}

\usepackage{algorithm}
\graphicspath{ {images/} }
\usepackage{algorithmicx}
\usepackage[noend]{algpseudocode}

\usepackage[utf8]{inputenc}
\usepackage[english]{babel}
\usepackage{lipsum}

\usepackage{amsthm}
 
\theoremstyle{definition}
\newtheorem*{definition}{Definition}

\begin{document}

\title{Signature-based IaaS Performance Change Detection}


\author{Sheik Mohammad Mostakim Fattah}
\affiliation{%
  \institution{School of Electrical Engineering, Computing and Mathematical Sciences, Curtin University}
  \state{WA}
  \postcode{6102}
  \country{Australia}}
\email{sheik.fattah@curtin.edu.au}

\author{Athman Bouguettaya}
\affiliation{%
  \institution{School of Computer Science, University of Sydney}
  \state{NSW}
  \postcode{2006}
  \country{Australia}}
\email{athman.bouguettaya@sydney.edu.au}

\renewcommand{\shortauthors}{S. Fattah and A. Bouguettaya}

\begin{abstract}
We propose a novel change detection framework to identify changes in the long-term performance behavior of an IaaS service. An IaaS service's long-term performance behavior is represented by an IaaS performance signature. The proposed framework leverages time series similarity measures and a sliding window technique to detect changes in IaaS performance signatures. We introduce a new IaaS performance noise model that enables the proposed framework to distinguish between performance noise and actual changes in performance. The proposed framework utilizes a novel Signal-to-Noise Ratio (SNR) based approach to detect changes when prior knowledge about performance noise is available. A set of experiments is conducted using real-world datasets to demonstrate the effectiveness of the proposed change detection framework.
\end{abstract}



\maketitle

\section{Introduction}

Infrastructure-as-a-Service (IaaS) is a primary cloud service delivery model that offers various computational resources such as CPU, memory, and storage \cite{jula2014cloud}. Several cloud services are provisioned through IaaS models, such as Virtual Machines (VMs), Virtual Storage (VS), and Virtual Private Networks (VPNs). IaaS models provide an easier and more cost-effective way to manage an organization's in-house IT infrastructure in the cloud \cite{iosup2011performance}. There are two primary subscription models for IaaS cloud services: a) pay-as-you-go and b) reservation. Reservation-based subscriptions are typically offered for a long-term period, such as 1 to 3 years. Large business organizations typically utilize IaaS cloud services on a \textit{long-term} basis \cite{van2014optimizing}. Most IaaS providers usually promote long-term subscriptions by offering significant discounts. For example, Amazon provides up to 72\% discounts on long-term subscriptions\footnote{https://aws.amazon.com/ec2/pricing/reserved-instances/}.  We consider any reservation-based subscription that is more than a month as a long-term subscription. 

Long-term IaaS cloud service \textit{selection} is a topical research area in cloud computing \cite{liu2015qos}. Subscribing to an IaaS service for a long period is an important \textit{business decision} for many organizations due to economic reasons \cite{mazzucco2011reserved}. The performance of IaaS services is one of the most important criteria for long-term selection \cite{iosup2014iaas}. Selecting a service that may exhibit poor performance in the future may cause a significant \textit{loss of revenue} for a business organization. IaaS performance is usually expressed in terms of Quality of Service (QoS) attributes such as CPU execution time, network throughput, and memory speed. QoS attributes are utilized to select the best service from a large number of functionally similar services \cite{zheng2013qos}.


 Knowledge of the IaaS services' performance is essential to determine which services best fit the consumers' required QoS  \cite{chaisiri2012optimization}. However, IaaS providers typically reveal very \textit{limited} performance information in their advertisements due to \textit{market competition} and \textit{business secrecy} \cite{wenmin2011history}. For example, most IaaS advertisements do not contain information about actual vCPU (virtual CPU) speed, memory bandwidth, or VM startup time. Moreover, the performance of a VM may fluctuate over time due to the dynamic nature of the cloud environment\cite{schad2010runtime}. As a result, advertised performance information may not reflect the true service performance for a particular provisioning time. For instance, a consumer may want to utilize some VMs in December, whereas the advertised performance is measured in June. In such a case, the advertised information is not useful for the selection in December. Additionally, the advertised performance information may not be \textit{helpful} to understand service performance due to the lack of detailed information \cite{fattah2020long}. For instance, Amazon does not provide detailed information about its different types of virtual CPUs (vCUPs). According to Amazon EC2 advertisements\footnote{https://aws.amazon.com/ec2/instance-types/}, each vCPU is either a thread of an Intel Xeon core, or a AWS Graviton processor. Estimating the performance accurately of a vCPU is difficult from such information. The lack of detailed and complete performance information makes the long-term selection challenging.  

Most IaaS providers offer \textit{free short-term trials} and encourage potential consumers to test their application workload in the cloud. Effective utilization of these free trials provides an effective alternative to deal with the limited performance information for the long-term selection \cite{fattah2020event}. However, free trial experiences do not provide adequate information to make an informed long-term selection. The key reason is that the performance of IaaS services may vary over time due to the multi-tenant nature of the cloud \cite{wang2018testing}. The observed performance in a trial may depend on the time of the trial. Therefore, making a long-term commitment based on only short trials does not always lead to the best service selection \cite{fattah2020event}.

\textit{IaaS performance signatures} offer an effective way to deal with the service performance variability for the long-term selection \cite{cherkasova2008anomaly,fattah2020icws}. An IaaS performance signature represents the \textit{expected} performance behavior of an IaaS service over a long-term period. For instance, a signature of a VM may indicate that its response time is expected to increase by 10\% in January than the response time in December. A consumer's trial experience of a service and its corresponding signature are utilized together to make an informed long-term selection in our previous work \cite{fattah2020icws}. \textit{We assume that a trusted third party collects performance data from free trial users to generate the signature of a service that represents its long-term performance behavior.} Geekbench\footnote{https://www.geekbench.com/} is an example of such a trusted third party where free trial users share their experience about different products or services. However, the service performance behavior may change over a period of time due to several reasons. For instance, a provider might upgrade its infrastructure or change its multi-tenant management strategy resulting in a change in service performance \cite{leitner2016patterns}. In such a case, the IaaS performance signature may need to be updated to be representative of the new performance profile of the service. It is therefore important to detect the change in IaaS performance \textit{as early as} possible to make sure its signature reflects the \textit{current} performance behavior of the service. \textit{We focus on the detection of long-term changes in IaaS performance behavior.} We represent the long-term performance behavior of an IaaS service using its performance signature. \textit{Therefore, the IaaS performance change detection problem is transformed into an IaaS signature change detection problem in this work.}

\textit{We identify two key challenges in IaaS performance change detection in the context of IaaS signatures}. The first challenge is determining the time when the signature needs to be re-evaluated for change detection. A change in the service performance behavior may occur at any time. The challenge is to accurately identify change points in time where there is a high likelihood of performance change occurrence. This is also known as the \textit{Change Point Detection} problem \cite{aminikhanghahi2017survey}. The second challenge is the ability to differentiate between the \textit{noise} and \textit{actual changes} in IaaS performance. Noise refers to the irregular or anomalous performance behavior that may not necessarily indicate long-term performance changes \cite{moens2019learning}. For instance, a major failure of computing infrastructure may negatively impact the performance of an IaaS service at a point in time without necessarily indicating a long-term change in the performance behavior. Given the multi-tenant and dynamic nature of the cloud, noise in IaaS performance is very common \cite{doelitzscher2013anomaly}. Therefore, the challenge is to identify performance noise to detect actual performance changes accurately. 

Detecting  changes  in  long-term  IaaS  performance  is important  as  it  will  help  new  consumers  to  select  the  best services according to their long-term QoS requirements. \textit{To the best of our knowledge, most existing research has not given enough attention to the long-term IaaS performance change detection problem \cite{fattah2019long}.} We propose an IaaS performance change detection framework in our previous work \cite{fattah2020event} that utilizes an Event-Condition-Action (ECA) model to detect changes in IaaS performance. The proposed work did not consider noise in IaaS performance during the change detection, which may substantially impact the change detection's accuracy. \textit{The focus of this paper is to distinguish the true changes in IaaS performance from the changes caused by performance noise}.

Noise in signal processing typically refers to the unwanted disturbance in electrical signals \cite{whalen2013detection}. Noise in different domains, such as signal processing and networking, is well-defined.  To the best of our knowledge, there is no performance noise model in the context of long-term IaaS performance. \textit{We propose a novel performance noise model and leverage existing change detection techniques to identify noise and changes in IaaS performance.} In particular, we define three types of IaaS performance noises in this work, i.e., spikes, attenuation, and distortion. We utilize the concept of IaaS performance noise to detect changes in IaaS performance signatures. We propose a novel change detection framework that identifies noises and true changes in IaaS performance signatures. The proposed framework consists of two change detection approaches. The first approach assumes that there is no prior knowledge about the nature of IaaS performance noise. The first approach therefore utilizes time series similarity measure techniques and a sliding window approach to identify different types of noise in IaaS performance signatures. The second approach assumes there is prior knowledge about the noise which is obtained from trial users' experience. In this case, we introduce a \textit{Signal-to-Noise Ratio}  (SNR)-based approach to detect changes in IaaS performance signatures. The key contributions of this work are summarized as follows: 

\begin{itemize}
    \item A novel IaaS performance change detection framework that utilizes the experience of free trial users and existing IaaS signatures to detect long-term changes in IaaS performance behavior.
    \item An IaaS performance noise model that represents different types of noise in IaaS performance signatures.
    \item A sliding window-based change detection technique to detect changes in IaaS performance signatures when there is no prior knowledge about the performance noise.
    \item A Signal-to-Noise Ratio based approach to detect changes in IaaS performance signatures when prior knowledge about performance noise is available.

\end{itemize}

\section{Background}

\begin{figure*}
    \centering
    \includegraphics[width=.8\textwidth]{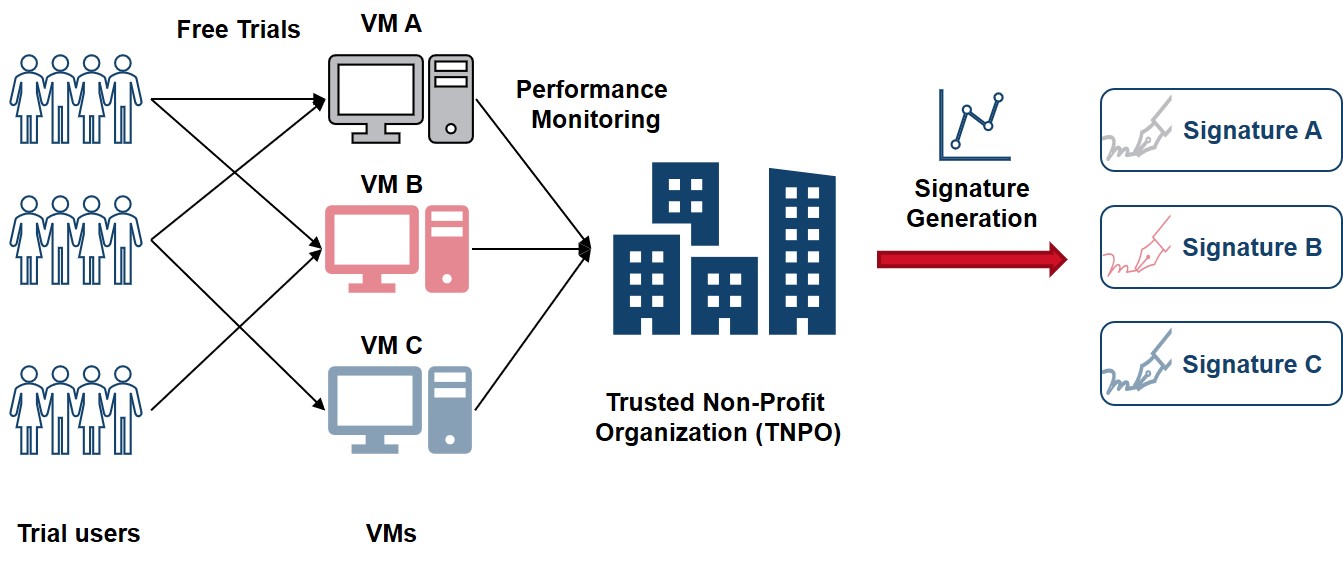}
    \caption{IaaS performance signatures generation}
    \label{fig:tnpo}
\end{figure*}

We discuss the representation IaaS performance signatures \cite{fattah2020event}. The word ``signature'' typically refers to the unique characteristics or behavior of an object, entity, or piece of information. The concept of the signature has been widely utilized in a number of domains such as cryptography, security, computing, and mathematics. For instance, performance signatures of various application are used to estimate resource capacity and detect performance anomalies \cite{mi2008analysis}. Intrusion Detection Systems (IDS) leverage signature-based malware detection techniques to enable quick detection of security threats. Digital signatures are commonly used to verify the authenticity and integrity of digital messages or documents. A checksum is also a form signature that is utilized to verify data integrity. Checksums are usually represented by a small-sized datum derived from block data to detect errors in data transmission. 



\subsection{IaaS Performance Signatures}

We represent the signature of an IaaS service based on its \textit{relative} performance changes over time, i.e., the amount of a service's performance may increase or decrease in one time compared to another time. For example, a VM's performance signature represents that its response time is expected to rise by 5\% on Sunday than Thursday. The performance signature of a VM does not represents the actual service performance. Therefore, a consumer is unable to select a service based on only its signature. Instead, the consumer needs to perform the trial with its application workloads and utilize its experience and the performance signature of the service to estimate the long-term performance \cite{fattah2020icws}. A signature is able to capture both weekly and annual performance variability since it is represented as related performance over time using a set of time series. As a result, it captures both seasonal effects, and trends in the performance behavior of a service. 

\begin{definition} {IaaS Performance Signature:} The performance signature of an IaaS service is the temporal representation of relative performance changes of an IaaS service over a long period.

\end{definition}

The signature is expressed using a set of QoS parameters that are typically the most important QoS parameters to measure the performance of a particular type of IaaS service \cite{fattah2020icws}. These parameters can also be considered as most relevant attributes of the service. For example, CPU speed and throughput are the key QoS attributes for vCPUs.

The performance signature of a service is represented as $S=\{S_1, S_2,...S_n\}$, where $n$ is the total number of relevant QoS parameters of the service. Each $S_i$ corresponds to a QoS parameter. Each $S_i$ represents a time series for $t$ period. The time series is represented as $S_i = \{s_{i1},s_{i2},......s_{it}\}$. Here, $s_{it}$ is the relative performance of the service at the time $t$. The following representation is used to represent the performance signature of a service:

\begin{gather}
 S =
  \begin{bmatrix}
   s_{11} & s_{12} & .. & s_{1t} \\
   s_{21} & s_{22} & .. & s_{2t}  \\
  s_{31} & s_{13} & .. & s_{3t}  \\
   .. & .. & ... \\
   s_{n1} & s_{n2} & .. & s_{nt}  \\
   \end{bmatrix}
   \label{eqn:signature}
\end{gather}

\normalsize where each row represents to the relative performance of $Q_i$ and each column denotes the timestamp $t$.

\begin{figure}

    \centering
    \includegraphics[width=.8\textwidth]{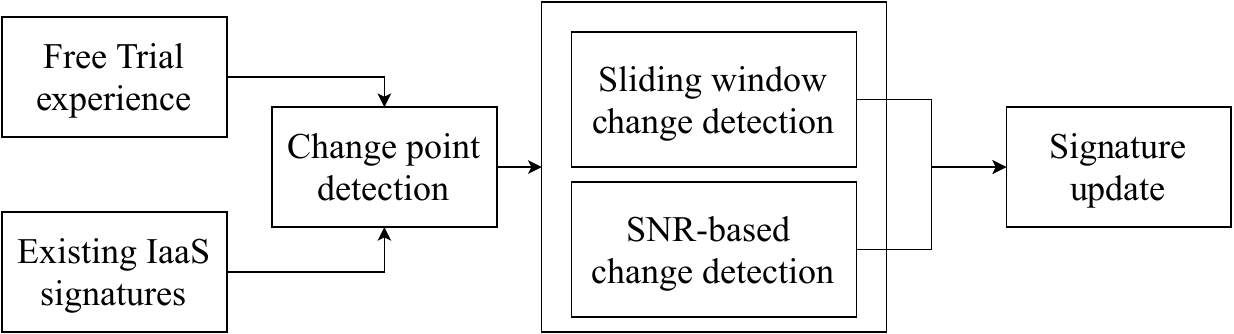}
    \caption{IaaS performance change detection framework}
    \label{fig:frame}
\end{figure}

\subsection{Signature Generation}

\label{sec:sig}

Trial users are typically reluctant to publish their data due to privacy, security, and the conflict of interests with the service providers \cite{zhu2015privacy}. However, they might share the data for a limited period to a third party (e.g., Trusted Non-profit organization) to help the community in making informed selection \cite{van2012trusted}. We often see examples of such third parties in public sectors where privacy-sensitive information is required to be shared to improve service delivery. For example, Cancer research centers usually collect information about patients to improve medical services. Such third parties or Trusted Non-Profit Organizations (TNPOs) typically perform data integration and distribution of collective knowledge. A closely related example can be found in the Geekbench website where trial experience of users is shared publicly and utilized by others to make an informed decision. These data however are generated based on geekbench tool which performs stress testing based on a few types of workloads on various platforms. To the best of our knowledge, we are the first to propose such use of trial data to help cloud consumers in long-term decision making. 

We assume that the trial users share their experience with a TNPO for a period of time. The TNPO analyze the trial experience and produces the performance signatures of the services. Once the signature is created, the TNPO deletes individual data. Fig. \ref{fig:tnpo} describes the proposed scenario in this work. There are three VMs offered by three IaaS providers. We assume that these VMs have almost identical virtual configuration. Free trial users utilize these VMs based on their application and share their trial experience with the TNPO. The TNPO creates signature for each VM to capture their relative performance variability for a long-term period. We assume that according to the contract between the trial users and the TNPO, the TNPO will delete the data of the individual trial experience. The generated performance signature represents the long-term performance behavior of a VM.

The signature needs to be generated in a way where trial experience of individual users can not be derived. This will ensure that the trial users' privacy will be protected. On one hand, the performance signatures is required to create in such a way that would require to store less information about individual experience. On the other hand, it has to be useful enough to help a new consumer in making long-term selection. Therefore, the signature is created using a \textit{normalized averaging} technique \cite{fattah2020icws}.

Let assume that there are $k$ trial users who shared their trial experience for a set of QoS parameters $Q={Q_1, Q_2,....Q_k}$ for $T$ period. $Q_{k}$ is the observed performance by $kth$ user for $Q$ QoS parameter. $Q_{k}$ is represented as a time series where $Q_{k}=\{q_{1k}, q_{2k},.., q_{tk}\}$. To obtain the signature for $Q$, the normalized averaging technique is applied as follows:

\begin{enumerate}[itemsep=0ex, leftmargin=*]
    
    \item For each QoS parameter $Q$, $k$ trial users' observed performance is gathered over time $T$.
    
    \item The average observed performance is measured for $k$ users at each timestamp $t \in T$. Let us denote the average performance of $Q$ as $\overline{Q_{k}}$.
    
    \item For each QoS parameter, $\overline{Q_{k}}$ is normalized using its standard deviation. The set of normalized QoS time series is the performance signature ($S$) of a service  for a period $T$. 
\end{enumerate}

In Equation \ref{eqn:signature}, each value of $s_{nt}$ denotes the relative performance at timestamp $t$ compare to any other timestamp $t'$ in the same row. There are two benefits of this representation. First, it makes the use of signature easier. If we can obtain a user's trial experience of a certain time, the signature can be used to derive performance in any other time. The second benefit is that we do not need to store detailed information about individual users. The signature represents significant changes in service performance behavior for a long-term period. As a result, the performance signature of a service should be reflected in the observed performance of most trial users.


\section{Proposed Change Detection Framework}

We discuss the proposed change detection framework in this section. IaaS performance change detection typically consists of two parts: a) change point detection (CPD), and b) change detection. The CPD process identifies the points in time when a change in performance may occur. Once a change point is identified, the change detection process recomputes the signature and evaluates whether the signature needs to be updated. The signature is then updated based on the evaluation of the change detection process. \textit{As our main focus is on the change detection process, we assume that the CPD process will be performed using the proposed approach in our previous work on change detection \cite{fattah2020event}.} The reason for including the CPD process is to provide a holistic view of the change detection process. Fig. \ref{fig:frame} shows the proposed change detection framework. The framework takes to input the free trial experiences of potential consumers and existing IaaS signatures. The change points are detected based on comparing the free trial experiences and existing IaaS signatures. Once a change point is identified, the change detection module applies two change detection approaches. The first approach assumes there is no prior knowledge about IaaS performance. This approach relies on time series similarity measure techniques and a sliding window approach. The second approach assumes that we have prior knowledge about the performance noise. The performance change is detected using prior knowledge about performance noise. In this approach, we utilize Signal-to-Noise ratio to differentiate between noise and change in IaaS performance. In the following subsections, we provide a brief overview of the change point detection and the change detection approaches.

\subsection{Change Point Detection}

Change point detection is formally known as the problem of finding abrupt changes in data when a property of a system of an entity changes. Change point detection is a pre-requisite of change detection, i.e., identifying the points in time when the likelihood of change occurrence is high. An ECA-based model is proposed in \cite{fattah2020event}. The ECA model is a simple yet powerful tool extensively used in databases, cognitive computing, and the semantic web. In the ECA model, when an \textit{event} is detected, a \textit{condition} is checked, and a resulting \textit{action} is executed \cite{liu2009encyclopedia}. The ECA model is especially useful when an action needs to be performed based on a condition that needs to be satisfied. According to the ECA model, an event determines when to trigger an action, the condition defines how to evaluate the event, and the action sets the execution plan in response to the event. An event is typically a special indicator that informs a system that action may need to be performed. An example of events in security software could be deleting many files at once. The security software may start evaluating the event, i.e., the deletion of many files, to determine whether it results from a security attack or a user action.

Anomalous performance behavior is a potential indicator of IaaS performance change \cite{mi2008analysis}. \textit{An anomalous performance behavior is the \textit{deviation} from the expected performance of an IaaS service.} The expected performance is represented by its performance signature. Performance anomalies are typically common in cloud environments \cite{ibidunmoye2015performance}. Anomalies may occur due to unexpected events faced by the IaaS provider, such as a sudden increase in the workload of the physical system, power failure, or natural disasters. As a result, experiencing performance anomalies in the free trial period may be normal in the cloud. However, the frequent occurrence of performance anomalies in the free trial period may indicate changes in IaaS performance, thus requiring a re-evaluation of the existing signature \cite{mi2008analysis}. Therefore, the event for the IaaS performance change point detection is defined as follows:

\noindent {\textbf{Event}:} An event is the frequent occurrence of performance anomalies that are experienced by the free trial users within a fixed period of time.

The frequency is initially defined as an arbitrary number or threshold $f$ which can be adjusted using a feedback loop from change detection outcome to change point detection. Once an event is detected, it needs to be evaluated to detect whether the signature has been changed. If the event satisfies the condition, the signature needs to be updated. The condition and action are defined as follows:

\noindent {\textbf{Condition}:} The condition is the process of testing an event to ascertain changes in IaaS performance.

\noindent {\textbf{Action}:}
The action is the process of updating the present an IaaS performance signature to reflect the changes in the IaaS performance.

The above three definitions are utilized as the basis for the ECA approach.

\subsection{ECA-based Change Point Detection}

\begin{figure}

    \centering
    \includegraphics[width=.8\textwidth]{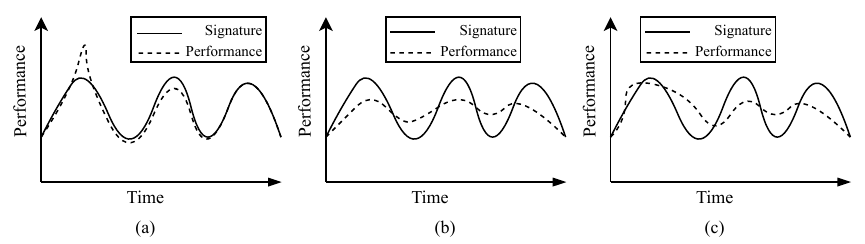}
    \caption{IaaS performance noise (a) Spike, (b) Attenuation, and (c) Distortion}
    \label{fig:noise}
\end{figure}

We utilize the free trial experience and the existing IaaS performance signatures to detect performance anomalies. The events are detected based on the performance anomalies. First, we measure the similarity between the trial experience of a consumer and the signatures to detect performance anomalies \cite{ibidunmoye2015performance}. When a user's trial experience is similar to the current signature, the signature is considered to be representative of the expected service performance. When the trial experience does not exhibit similar performance behavior as represented by its signature, we consider it as an anomalous performance behavior. The signature represents the relative performance behavior as a time series. As a result, the shape of the time series needs to be considered to measure the similarity rather than the value of each data point in the signature time series. There are numerous approaches in the existing literature to measure time series similarity based on the shape such as Pearson Correlation Coefficients (PCC), Euclidean Distance (ED), Cosine Similarity (CS), Symbolic Aggregate Approximation (SAX), and Dynamic Time Warping (DTW) . Each of these methods may be applied to measure the similarity between the trial experience and the signatures. We briefly discuss how to apply the PCC, CS, and ED for the similarity measure to detect performance anomalies in the context of IaaS performance signatures.

Let us denote the trial experience of a user by $E_Q$ where $E_Q$ denotes the performance of an IaaS services in the free trial period  $T_f$. Here, $T_f << T$, i.e., the free trial period is significantly less than the required provisioning time $T$. We represent $E_Q$ as a time series $E_Q = \{q_1, q_2, ...q_{n}\}$ where $n$ is the number of timestamps in $t$. $E_Q$ needs to be normalized before measuring the similarity with an IaaS performance signature. Let $E'_Q$ is the normalized trial performance where the normalization is performed based on its standard deviation. We denote $E'_Q$ as $E'_Q = \{q'_1, q'_2, ...q'_{n}\}$. Let the signature of an IaaS service for the trial period $T_f$ is $S_Q$ for the QoS attribute Q where $S_Q = \{s_1, s_2, s_3,...s_n\}$. The similarity between the normalized trial experience $(E'_Q)$ and the signature of a service during the trial period $(S_Q)$ using the Euclidean distance $(S(E'_Q, S_Q)^{ED})$ is computed by the following equation:

\begin{equation}
    S(E'_Q, S_Q)^{ED} = \sqrt{\sum_{t=1}^n (s_t - q'_t) }
\end{equation}

\normalsize

Similarly, the similarity measure using the Pearson Correlation Coefficients is computed using the following equation: 

\begin{equation}
    S(E_Q^N, S_Q)^{PCC} = \frac{\sum_{t=1}^n (s_t - \Bar{s}) (q'_t - \Bar{q'}) }{\sqrt{ (s_t - \Bar{s})^2 } \sqrt{(q'_t - \Bar{q})^2}}
\end{equation}

\normalsize

where $\Bar{q'}$ and $\Bar{s}$ is the mean value of $q'$ and $s$ within the trial period $T_f$. The cosine similarity of the trial experience is measured by the following equation: 

\begin{equation}
    S(E_Q^N, S_Q)^{CS} = \cos{\theta} = \frac{\sum_{t=1}^n s_t  q'_t }{ \sqrt{\sum_{t=1}^n (s_i)^2 } \sqrt{\sum_{t=1}^n (q')^2_i}  }
\end{equation}

\normalsize

Each of the above equations provides us with a similarity value between the trial experience and the corresponding IaaS performance signature. In the case of the euclidean distance, the lower the distance is the higher the similarity. 

A similarity threshold needs to be defined to determine how much deviation of the performance from the signature should be considered as the performance anomaly. We define a similarity threshold $S_{thresh}$ for each technique. The threshold is used to distinguish between the normal performance behavior and performance anomalies. The initial threshold is defined during the signature generation process based on the experience of the past trial users' experience. Let us assume that there are $N$ number of past trial users. The experience of the past trial users is denoted by $E_P = \{E_1, E_2, ...E_N\} $. The initial similarity threshold $T_S$ for anomaly detection is defined as follows:

\begin{equation}
    T_S = \min_{i=1}^N{S(E_i, S_Q)}^{M}
\end{equation}

\normalsize

where $M$ denotes the similarity measure method, i.e., PCC, ED, or CS. The threshold for different similarity measure technique can be different. When a new user performs trial if the user's observed performance has a similarity lower than the $T_S$, we consider it as anomalous performance behavior of the service. The value of the similarity threshold $S_{thresh}$ can be updated during the training of the proposed ECA model.

The event for signature change detection is defined as the frequent occurrence of performance anomalies within a fixed period of time as mentioned earlier. We define an anomaly threshold for the event detection and denote it as $F_{thresh}$ which represents the minimum number of occurrences of the performance anomalies within a period of time $T_f$. The value of $T_f$ is the length of the free trial period. We assume that each provider offers the same length of free trial without the loss of generality. The value of $F_{thresh}$ can be initially defined as the number of past trial users within each $T_f$ period who have the minimum similarity between their experience and the corresponding signature. For example, if there are 5 past trial users who have the minimum similarity $T_S$ with the present signature, then $F_{thresh}$ is initialized as $5$. In such a case, the number of past trial users that have the minimum similarity during the signature generation process is considered as the usual number of performance anomalies within the $T_f$ period. When the number of performance anomalies crosses $F_{thresh}$, we consider it as an event that needs to be evaluated for signature change detection. We update the value of $F_{thresh}$ over time to detect the signature change effectively.

\subsection{IaaS Performance Noise}

A change point indicates that a change may have occurred in IaaS performance behavior. After a change point has been detected, the existing signature needs to be tested to ascertain whether a revaluation of the signature is required. The main challenge in ascertaining is to differentiate between the noise and change in performance. To the best of our knowledge, there is no existing model that would help us to identify the performance noise. It is therefore very difficult to find empirical evidence of performance noise. Since there is no theoretical model, we followed a trial and error-based approach where we proposed a new performance model based on the signal processing domain and evaluated it through the experiments. The proposed noise model is corrected over time based on the evaluation. \textit{We define noise as the irregular or abnormal performance behavior in the performance of an IaaS service with respect to the corresponding IaaS signature.} We identify the following three types of noise that may appear in IaaS performance as shown in Fig. \ref{fig:noise}. It is possible that there are other types of noises. However, it can be investigated in future work.

\begin{enumerate}[itemsep=0ex, leftmargin=*]

    \item \textbf{Spikes}: Spikes in IaaS performance typically originate from various uncertain and sudden changes in performance (Fig. \ref{fig:noise}(a)). For instance, an IaaS provider may face a major power failure of its infrastructure. These types of noise typically stay for a short period. The performance becomes normal once the provider resolves the issue.
    
    \item \textbf{Attenuation}: The second type of noise is caused by attenuation. Attenuation in a signal usually refers to the loss of power in signal strength. In the context of the IaaS signature, it also has a similar effect (Fig. \ref{fig:noise}(b)). The performance of an IaaS service may not increase or decrease as expected according to the signature. However, the performance has a similar shape to its signature. Attenuation may occur due to the effect of multi-tenancy in the cloud. 
    \item \textbf{Distortion}: Distortion refers to random noise in performance (Fig. \ref{fig:noise}(c)). These types of noise are very irregular and do not follow any particular distribution. Distortion may be originated from multi-tenancy, scheduled maintenance, security attacks, sudden hazards, and so on. Differentiating these types of noises from performance change is most challenging. 
\end{enumerate}

First, we develop a change detection approach that applies a sliding window-based detection technique that identifies the noises in IaaS performance. In this case, we assume that we do not have any historical knowledge about performance noise. Next, we develop an SNR-based approach to improve the performance of the change detection process. In the second case, we assume that we have prior knowledge about the nature of noise. In the following sections, we describe these two approaches.

\subsection{Sliding Window Change Detection}

Once a change point has been detected, the signature is recomputed based on the experience of current trial users. The existing signature and the recomputed signature need to be compared to detect changes in performance. Both signatures are represented as time series. Therefore, we may utilize existing time series similarity measure techniques to compute the similarity between two signatures. However, most of these techniques do not consider noise in time series during the similarity measure. Therefore, we develop a sliding window-based noise detection technique to differentiate between noise and changes in IaaS performance.

The sliding window-based approach leverages existing time series similarity measure techniques. In particular, we will use PCC and RMSE (Root Mean Squared Error) to measure similarities between two signatures based on shape and distance. We utilize these two similarity measure techniques as they are effective, easy to implement, and most commonly used techniques. We start with the assumption that there is no noise in the IaaS performance. Therefore, we measure the similarity based on PCC and RMSE distance between the existing signature and the recomputed signature. The PCC is used to compare the shape between two time series. The next step is to measure the distance between the two signatures using RMSE.

\begin{figure}

    \centering
    \includegraphics[width=.5\textwidth]{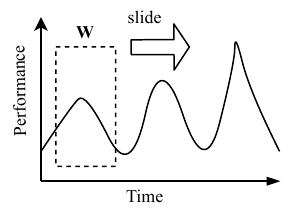}
    \caption{Sliding window approach to identify performance noise}
    \label{fig:slide}

\end{figure}






When the existing signature and the recomputed signature have high similarity and a low RMSE distance, we assume that the existing signature does not require re-evaluation. Therefore, the detected change point would be considered a false positive. We need to define two thresholds to determine whether there is a high similarity and low RMSE distance between these signatures. Therefore, we define the following two thresholds, $S^P$ and $S^R$  for the PCC and RMSE, respectively. Since there is no standard way to define these thresholds, we defined these thresholds iteratively to find the appropriate values during the training phase of the change detection. We choose 0.6 as the value of $S^P$ and .2 as the value of $S^R$.  If the PCC or RMSE is lower than these thresholds, we need to decide whether the existing signature needs to be updated. If we can identify the noise in the recomputed signature, and by removing the noise, the similarity increases considerably, we do not change the existing signature. If the shapes of the signatures are very similar, however, the RMSE distance is within an acceptable threshold $T^{D}$, we consider it as the effect of attenuation. Therefore, we do not change the existing signature in this case. If the PCC value is low and the RMSE distance is also low, that indicates the shape has been altered. It could have resulted from either performance change or noise. If it is because of noise, then the noise is either spike or distortion. We can detect spikes in IaaS performance using a sliding window-based approach, as shown in Fig. \ref{fig:slide}. In this approach, we start with an assumption that the lower value of the PCC originated from spikes. Therefore, if we can identify and remove spikes from the recomputed signature, the similarity score will increase. Therefore, we define a window $W$. The length of $W$ should be greater than on the size of the spike to identify the noise. Therefore, the length of $W$ should be defined based the size of the spike. For example, a spike in practice should last for a short period, e.g., 1 hour.  We scan the signature from beginning to end using the $W$ length. Each time we remove the $W$ portion of the signature and recompute the value of PCC. If the similarity score increases and reaches up to the threshold $S^P$, then we consider it noise. Therefore, we do not update the existing signature. We are unable to distinguish between the change and noise in this approach if the noise originated from distortion. In this case, we need to rely on past knowledge about the noise.

\subsection{SNR-based Change Detection}

In this section, we discuss the change detection approach with prior knowledge about performance noise. In this approach, we rely on historical information instead of time series similarity measures. As a result, we do not need to identify different types of noise separately. We leverage a noise detection technique from the signal processing domain to identify noise in IaaS performance. 

We assume that once a signature has been generated, the TNPO monitors the performance of free trial users for a period $T$ (e.g., 1 year). In this period, we assume that the IaaS signature of a service does not change considerably. Therefore, the TNPO monitors performance without any intervention to the existing signature to understand performance noise. The TNPO monitors the noise and measures the Signal to Noise Ratio (SNR) \cite{bruzzone2000automatic}. SNR is used to measure the level of the desired signal and the level of background noise. SNR is typically expressed as the ratio of signal power to the noise power, often expressed in decibels. A ratio higher than 1:1 (greater than 0 dB) indicates more signal than noise. In the context of change detection, we consider the existing IaaS signature as the desired signal and the IaaS performance noise as background noise. The signal to noise ratio is computed as follows: 

\begin{equation}
    SNR = \frac{E(S^2)}{E(N^2)}
\end{equation}

where $S$ is the signal and $N$ is the noise. $E$ refers to the expected value, i.e., in this case, mean square of $S$ and $N$. When the noise has an expected value of zero, then the denominator is its variance instead of the mean. We assume that the TNPO divides the $T$ period into $d$ number of segments. The value of $d$ can be adjusted based on understanding the noise in the different seasonal periods, i.e., daily, weekly, and monthly. The TNPO requires recomputing the signature in these $d$ periods to compute the noise. The noise ($N$) can be computed by measuring the difference between the existing signature and the recomputed signature using the following equation:

\begin{equation}
    N = S - S^r
\end{equation}

where $S$ is the existing signature and $S^r$ is the recomputed signature in the noise learning period. For each $d$ period, we measure the SNR value. Once the monitoring period is finished, we start the change detection process. Once we detect a change point, we recompute the signature based on the experience of current users. We then compute the SNR value of the current signature compared to the recomputed signature. If the current SNR value is less than the existing SNR value, we consider that a change has occurred.

\section{Experiments}

In this section, we discuss a set of experiment results to evaluate the effectiveness and the efficiency of the proposed sliding window and SNR-based approach. We compare the proposed approaches with each other and with the existing IaaS performance change detection approach, i.e., cumulative sum control chart (CUSUM) proposed in \cite{fattah2020event}.

\subsection{Experimental Settings}

\begin{figure}
 
 \centering
 
 \includegraphics[width=0.5\textwidth]{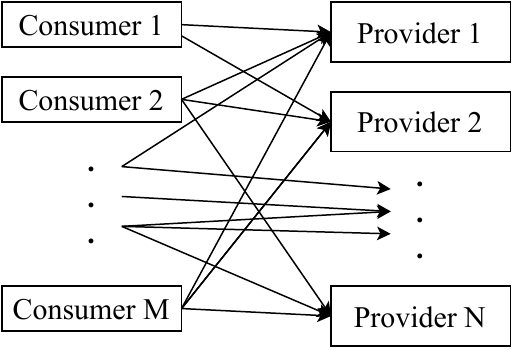}
 
    \caption{Environment of the experiment}
    
    \label{fig:exp_scene}
\end{figure}

To conduct the experiment, it is necessary to have the trial experiences of a set of consumers over a long period. Each consumer runs their workloads on a variety of IaaS services to select the best service based on their requirements. Fig. \ref{fig:exp_scene} illustrates an environment where consumers perform free trials on a set of IaaS providers. We assume that each provider represents an IaaS service. The experience of free trial users needs to be shared with a TNPO who generates the performance signature of each service. Therefore, we need a set of trial users' workloads, corresponding IaaS performance of a set of IaaS services to create performance signatures. In addition, we require information when a provider's performance behavior changes as ground truth to evaluate the proposed framework.

\subsection{Experiment Datasets}

We utilize the dataset used in our previous work \cite{fattah2020event} on the change detection. We provide a detailed description of the experiment datasets in the following subsections. 


\subsubsection{Trial Users' Workload Datasets}

Consumers perform trials based on their workloads. We obtained the workload data for trial from Eucalyptus IaaS cloud workload traces \cite{wolski2017qpred}. To the best of our knowledge, these traces are the best-fit for our experiment requirements. These traces represents real-world cloud environments \cite{wolski2017qpred,pucher2016using}. It contains workloads from a large company with 50,000 to 10,000 employees. The traces are anonymized and they represent six production systems that utilize the Eucalyptus IaaS cloud. The trace contains IaaS service level and physical level information. At the service level, it contains information about request timestamps, instance lifetime, instance requests, and timestamps. At the physical level, it contains information about the number of cloud nodes, CPU cores, and physical host occupancy. There are six traces in the Eucalyptus IaaS cloud traces. We select the trace ``D6traces'' as it is one of the largest traces that contain approximately workloads of 34 days of 31 nodes where each node contains 32 cores. Each node is utilized to represent the workload of a trial user. The trial user's workload contains nearly 65,000 timestamps. We apply piece-wise aggregate approximation (PAA) to obtain 360 timestamps where we assume each timestamp represents the average workload of a day of a trial user. The workload dataset's description is shown in Table \ref{tab:data1}.

\begin{table}

\caption{Workload Trace Description} \label{tab:data1}
\centering
\begin{tabular}{|l|r|}
 \hline
 Parameter & Value \\
 \hline
 Number of cloud nodes & 31 \\
 Number of CPU cores at each node & 32 \\
 Number of timestamps & 6486 \\
 Resource allocation unit & number of CPU cores \\
 Trace attributes & VM start, stop timestamps, \\
                 & VM resource requests, \\
                 & VM id, node id  \\
 \hline
 \end{tabular}

\end{table}

\subsubsection{IaaS Performance Dataset Generation}

\begin{figure}

 \centering
 
 \includegraphics[width=0.6\textwidth]{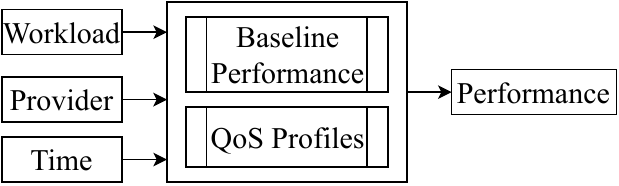}

    \caption{Data generation framework}
   
    \label{fig:data_gen}
\end{figure}

It is very challenging to find a performance dataset for long-term workload traces. As a result, we decided to synthesize the performance dataset for the workload traces we've chosen. We build a framework for generating QoS performance of a set of IaaS services based on the given workload and time period. The performance data generation framework is shown in Fig. \ref{fig:data_gen}. It takes a trial user's workload, a provider, and time as inputs. Based on the given inputs, it generates the corresponding QoS performance of the service. The framework has two components which are the baseline performance and QoS profiles.

\begin{itemize}
    \item Baseline Performance: The baseline performance represents a workload-performance map that does not contain any performance variability. It gives an expected performance for a given workload. This performance is then transformed based on the QoS profile of an IaaS service to reflect realistic cloud performance. The baseline components primarily map an given workload to an initial performance value. The generated performance value is later transformed using the QoS profiles to obtain the performance of a given workload for a service at a particular time.  We generate the baseline performance using the SPEC benchmark results published in 2016 \cite{baset2017spec}. We create a workload-QoS map based on the collected performance data from the benchmark result, which contains about 1,500 observations. Each unique workload request is mapped to a specific performance value based on the resource requirements. The request with the highest resource requirements is mapped to the lowest performance value, and the request with the lowest resource requirements is mapped to the highest performance value. The reason is that the performance of a VM typically degrades when it serves a resource-intensive workload. For example, if we increase the stress on the CPU, the overall throughput of a VM will degrade. If we decrease resource consumption, the performance will increase.

    \item QoS Profiles: We create five QoS profiles to simulate the long-term performance behavior of five IaaS services from five IaaS providers. Each QoS profile decides the performance of a service for a given workload in a particular timestamp. The QoS profile of each service is represented using a relative performance index which is utilized together with the baseline performance to generate the final performance value. The relative performance index has a workload map and a seasonal map. The workload map provides the performance of a service for a given workload compared to the baseline performance. The seasonal map decides the effect of time on the performance value compared to the baseline performance. 
    
    For each provider, we define a unique workload map. The workload map takes the given workload and the baseline performance to generate the expected performance of a provider. For example, if the given workload requests 60\% CPU, and the baseline performance is about 1.5K operations/sec, then the workload map of provider A generates 1.65K operations/sec, which is about 10\% more than the baseline performance. The workload map basically contains a provider-specific set of rules which produces expected performance based on the workload and baseline performance. Similarly, the seasonal map of a provider contains a set of rules that produces the final performance based on the timestamp and the expected performance. The seasonal map of a provider takes the time of the workload and the expected performance to generate the final performance. For instance, if the expected performance is about 1.65K operation/sec, and the time is December, then the final performance would be 1.68 operation/sec, which is about 1\% higher than the expected performance. 
    
    The workload map and the seasonal map are created manually for five providers based on real-world cloud characteristics as reported in \cite{iosup2011performance}. According to this study, cloud performance may vary considerably based on workload and time. It shows that some services exhibit daily and yearly seasonality. In the case of the seasonal map, we only consider yearly seasonality. To simulate the performance variability based on the workload, i.e., the same workload may get a different performance, we include a small noise to the final performance generation based on a uniform distribution. The noise was created using a uniform random generated between 0 to 1. The study also suggests that the performance of similar services from different providers may exhibit different performance behavior. Therefore, we made each provider’s QoS profile different. 
    

\end{itemize}

We create the signature of each provider using the signature generation approach described in Section \ref{sec:sig} based on the generated performance data of each provider. The experiment variables are shown in Table \ref{tab:data}. We conduct the experiments by changing the signatures to simulate the change in performance behavior of IaaS providers. We create some changed signatures by replacing a part of the original signature with different signature. The different signature is chosen randomly from one of the five providers. We then create some noisy signatures by adding noise to the original signature. We insert two types of noise spike and \textit{Additive white Gaussian noise} (AWGN), which mimics the effect of many random processes in nature. Since we have built the dataset for this evaluation, the result may be sensitive to the provided input. We therefore run the experiment 30 times and use a large number of signatures (6,000) to mitigate the effect of input variables. We have developed the experiment using Matlab on a computer with Intel Core i7 (2.80 GHz and 8Gb ram).

\begin{table}

\centering
\caption{Experiment Variables}\label{tab:data}
\begin{tabular}{|l|l|}
\hline
{\bfseries Variable Name} & {\bfseries Values}\\
\hline
Total provisioning period & {360}  days \\
Trial size & {30} days \\
Total number of IaaS performance signatures & {5} \\
Total number of noisy IaaS performance signatures & 3,000 \\
Total number of changed IaaS performance signatures & 3,000 \\
Similarity threshold $S^P$ & 0.60 \\
RMSE distance threshold $S^R$ & 0.20 \\
Sliding Window Size $W$ & 6 days \\ 
Spike size & 3 \\ 
Additive White Gaussian Noise & 20 SNR \\

Number of Simulation & 30 \\ 

\hline
\end{tabular}

\end{table}

\subsection{Evaluation of the Proposed Approaches}

\begin{figure}

    \centerline{
        \subfloat[]{\includegraphics[width=0.5\textwidth]{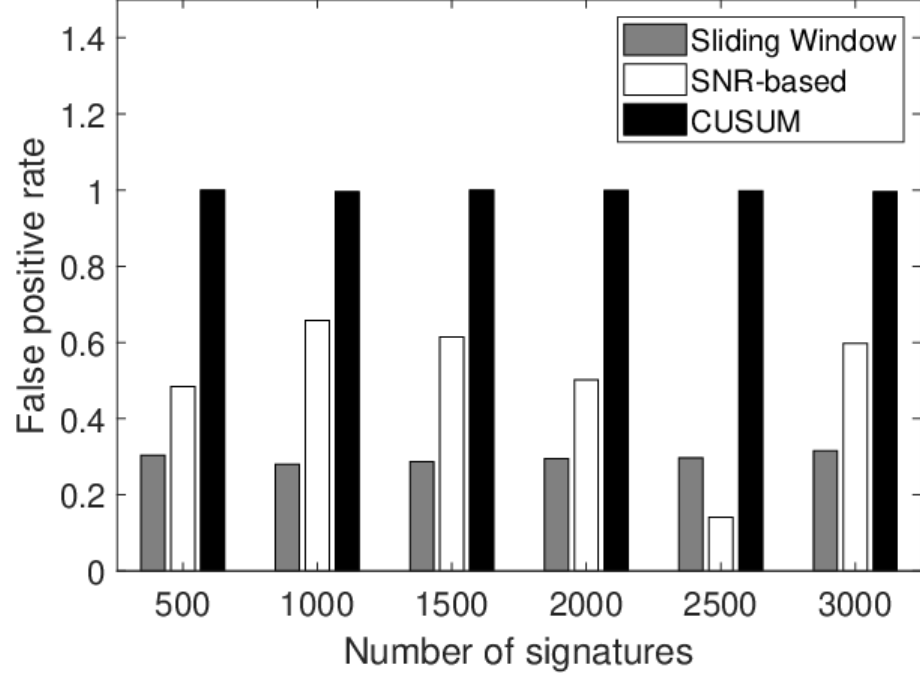} \label{fp_rate}}
        \hfil
        \subfloat[]{\includegraphics[width=0.5\textwidth]{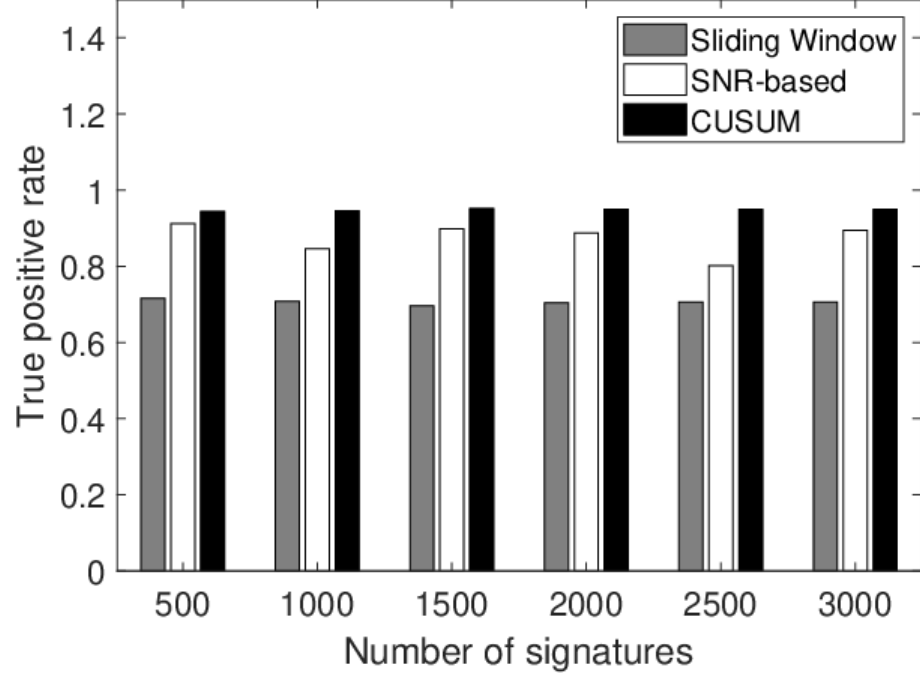} \label{tp_rate}}
    }
 
    \caption{ (a) False positive rate (b) True positive rate }
  \label{fig:exp1}

\end{figure}

The aim of the experiments is to evaluate the effectiveness and the efficiency of the proposed approach in terms of their ability to differentiate between noise and true performance changes. Therefore, we consider four attributes: 1) True Positives (indicates correct change detection) (TP), 2) False Positives (FP) (indicates incorrect change detection), 3) True negatives (indicates correct noise detection) (TN), and 4) False negatives (indicates incorrect noise detection). These metrics are widely used metrics to evaluate performance of change detection approaches. First, we look at the false positive rate and the true positive rate of the proposed approaches. The false positive rate (also known as the false alarm ratio) indicates the expectancy of the false positive ratio and is calculated as follows: 

\begin{equation}
    \text{False Positive Rate (FP rate)} = \frac{FP}{FP+TN}
\end{equation}

The lower value of false positive rate indicates better performance. Similarly, true positive rate (also known as recall or sensitivity) refers to the ability to identify correct changes and is calculated as follows:

\begin{equation}
    \text{True Positive Rate (TP rate)} = \frac{TP}{TP+FN}
\end{equation}

Fig. \ref{fig:exp1}(a) shows the FP rate of the three different change detection approaches. The experiment is run with five different sample sizes of IaaS signatures. $X$-axis indicates the sample size of each iteration. $Y$-axis indicates the false positive rates. Fig. \ref{fig:exp1}(a) shows that the sliding window approach has the lowest average FP rate compared to the other two approaches (around 0.3 in most cases). The SNR-based approach has a higher FP rate than the sliding window approach and a lower FP rate than the CUSUM approach. The CUSUM approach has the highest false positive rate. Similarly, Fig. \ref{fig:exp1}(b) shows the TP rate for the three approaches. The sliding window has the lowest TP rate, and the CUSUM has the highest TP rate. The SNR-based approach provides a TP rate better than the sliding window and lower than the CUSUM. These results indicate that the CUSUM is unable to distinguish between the noise and actual change in IaaS performance. Therefore, it considers all types of noise and change as true change. As a result, it has the highest FP rate and TP rate. The sliding window-based approach offers the lowest FP rate. However, its ability to detect true changes is also lower compared to the other approaches. The SNR-based approach exhibits a balanced FP and TP rate compared to the other two approaches. As a result, the SNR-based approach can be considered to have a better ability to distinguish between the noise and changes in IaaS performance. Now, we measure the accuracy and the f1 score of these approaches to further evaluate the results. Accuracy and F1 score are computed as follows: 

\begin{equation}
    \text{Accuracy} = \frac{TP + TN}{TP+FP+TN+FN}
\end{equation}

\begin{equation}
    \text{F1 score} = \frac{TP}{TP + \frac{1}{2}(FP+FN)}
\end{equation}

\begin{figure}
    \centerline{
        \subfloat[]{\includegraphics[width=0.5\textwidth]{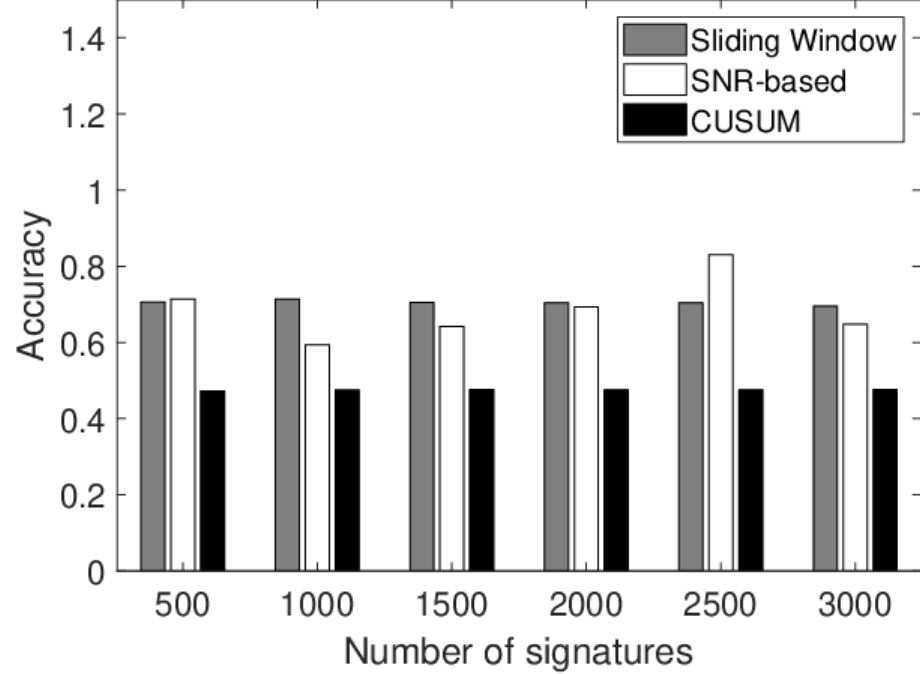} \label{ac}}
        \hfil
        \subfloat[]{\includegraphics[width=0.5\textwidth]{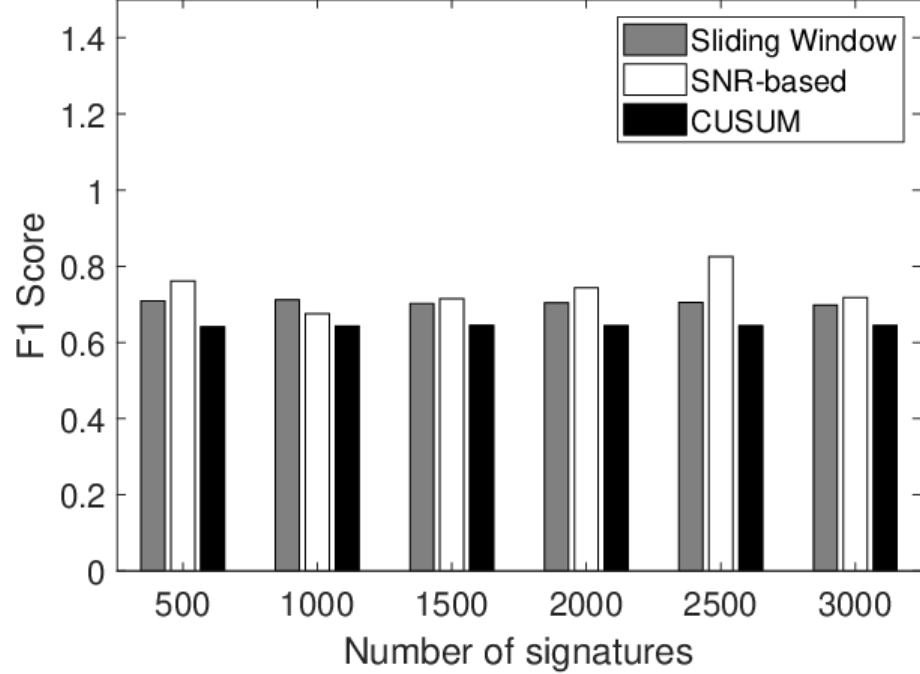} \label{f1}}
    }
 
    \caption{ (a) Accuracy (b) F1 Score }
  \label{fig:exp2}
\end{figure}

Accuracy refers to the ability to correctly detected changes compared to the total number of observations. Fig. \ref{fig:exp2}(a) shows the accuracy of the three approaches. It shows that the sliding window has the highest accuracy most of the time, and CUSUM has the lowest accuracy. The SNR-based approach has an accuracy higher than the CUSUM, and lower than the sliding window approach. From the accuracy results, the sliding window approach appears to be the best performing approach. However, the accuracy metric for change detection is not always suitable. Because the accuracy result can be biased if there is a large number of true negatives in the data. In such a case, the F1 score is a better indicator of performance measure. F1 score considers both false positives and false negatives into account. F1 is therefore a better indicator for uneven class distribution of data. Fig. \ref{fig:exp2}(b) shows the F1 score of the three approaches. The CUSUM has the lowest F1 score, and the SNR-based approach has the highest f1 score. The sliding window has a lower F1 score compared to the SNR-based approach. This indicates that the SNR-based approach has a better ability to distinguish between the noise and changes in IaaS performance compared to the other two approaches. This conclusion is the same as the conclusion we derive from the FP rate and the TP rate in Fig. \ref{fig:exp1}. 

\subsection{Sensitivity Analysis}

\begin{figure}

    \centerline{
        \subfloat[]{\includegraphics[width=0.5\textwidth]{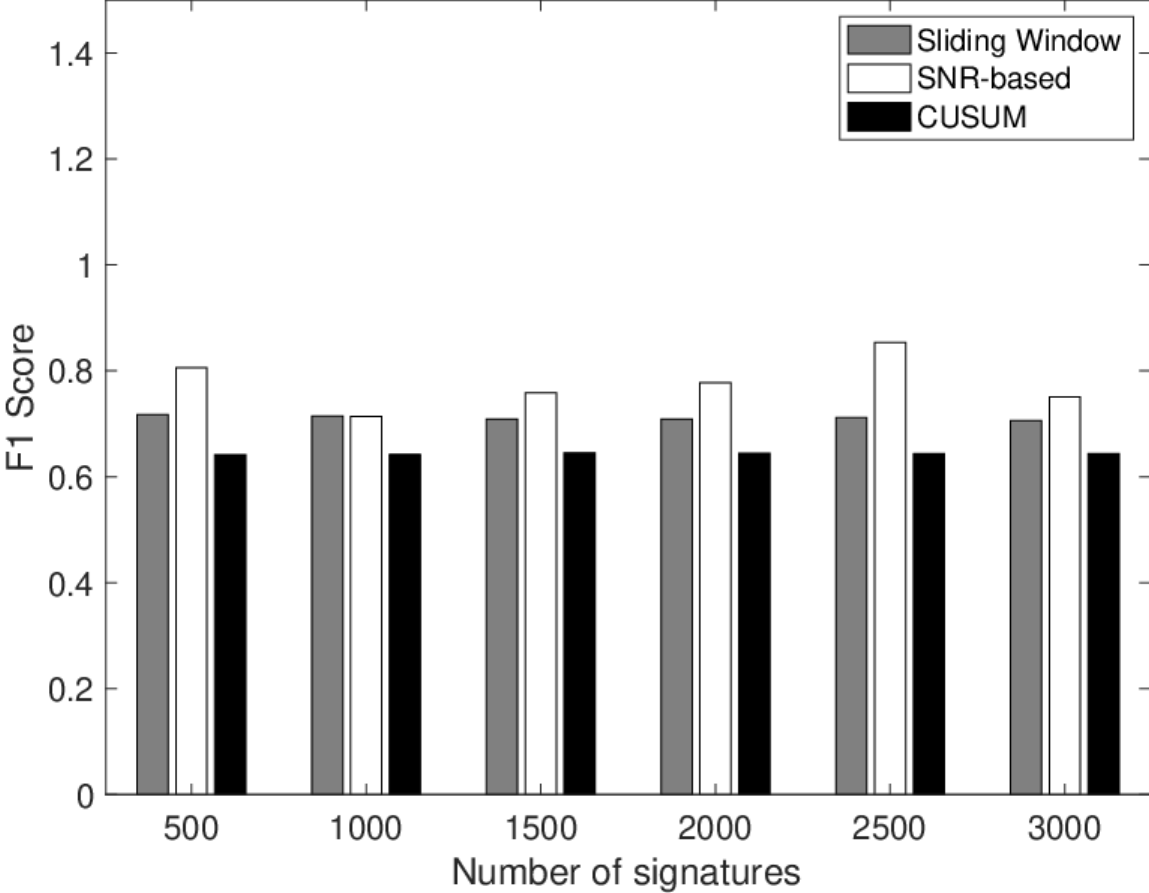}}
        \hfil
        \subfloat[]{\includegraphics[width=0.5\textwidth]{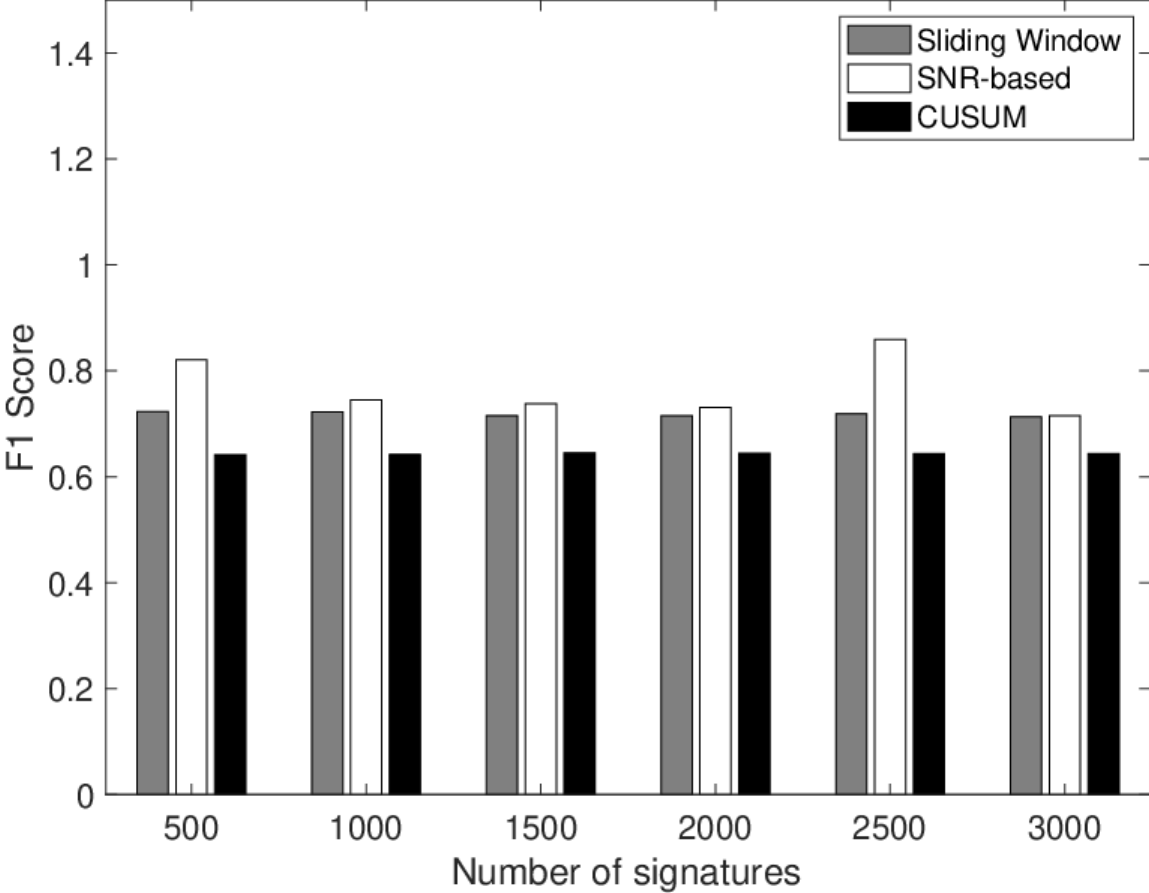}}
    }
 
    \caption{ Sensitivity analysis (a) 50\% reduced distortion (b) No distortion in the dataset }
  \label{fig:exp3}

\end{figure}

We have leveraged existing datasets and augmented them to conduct the experiment. To mitigate the effect of variable input on the output, we run our experiment multiple times and work with a large number of signatures, as shown in Table 2. It presents the average performance of our proposed approach. We consider the sliding window-based approach as an alternative to the SNR approach when there is no prior knowledge available about the nature of the noise. SNR-based approach supposed perform better in most cases, since it relies on prior knowledge. Therefore, even if the dataset has zero changes caused by distortion, the SNR-based approach will be able to detect changes properly for other types of noises. To provide empirical evidence, we conduct further experiment that shows that the SNR-based approach still provides a better F1-score than the SW-based approach as shown in Fig. \ref{fig:exp2} and Fig. \ref{fig:exp3}. In the experiment, we perform a sensitivity analysis by varying the number of signatures that has distortions. The initial dataset has 50\% noisy signatures (both distortion and spikes) and 50\% changed signature. In the next stage, we remove distortion from half of the noisy signatures. Finally, we completely remove distortion from all noisy signatures. The SNR-based approach performs better in all three cases.

\subsection{Discussion}

The experiment results show that the proposed sliding window-based approach and the SNR-based approach are able to detect noise and performance changes in IaaS signatures effectively. However, there are a few threats to validity of our study and evaluation. First, we mainly focus on three types of performance noises inspired from signal processing domain. However, there could be other types of noise that may impact the accuracy of the change detection. The future work of this study should investigate other types of noise and approaches in the context of IaaS cloud services. Second, the SNR-based approach is not applicable if there is no prior knowledge. In this case, we rely on the sliding window-based approach which is unable to detect the effect of distortion in the performance signature. Therefore, we need to investigate new approaches to identify distortion to detect changes without prior noise information. A key limitation of this work is the use of a short-term performance dataset as we are unable to find any suitable long-term performance dataset. Therefore, the accuracy of the proposed approach is mainly applicable to the utilized dataset. Another limitation of this work is the lack of measurement information about other attributes such as communication between two nodes in the Eucalyptus clusters. These limitations could be overcome by developing a testbed to generate long-term cloud workload-performance dataset. There are a few thresholds that are set during the experiment based on the trial since there is no standard way of setting these values in the case of noisy performance. The value of these thresholds may affect the performance. We have listed the values of these thresholds in Table \ref{tab:data}. In the future work of this study, a systematic approach can be developed to deal with similarity and distance thresholds. In several cases, we have used a random number generator from a uniform distribution. For example, during the changed signature generation we chose a random provider's signature to replace a part of the original signature. For simplification we have used such uniform distribution. In the future work, a more systematic approach could be developed for such random steps.

\section{Related Work}

IaaS performance is one of the key criteria  select the best IaaS cloud services for a consumer \cite{iosup2011performance,eismann2022case}. Therefore, the performance of IaaS cloud services has been studied in many studies \cite{iosup2014iaas,leitner2016patterns, wang2018testing,fattah2019long,hao2021empirical,he2021performance}. The performance of IaaS cloud services is typically measured for diverse applications based on short-trials in the cloud \cite{wang2018testing}. A set of experiments is conducted to study the performance of Amazon EC2 instances for service-oriented applications \cite{dejun2010ec2}. It runs benchmarks on VMs using a wide variety of workloads. The performance of the instances are analyzed based on their response time. An automated performance testing approach is proposed \cite{jayasinghe2012expertus} for IaaS cloud called the generator approach. The proposed approach leverages a template-drive method for IaaS cloud performance testing. An empirical study is carried out to study the VM startup times in public IaaS clouds in \cite{hao2021empirical}.  

An extensive study on the performance variability of Amazon EC2 is provided in \cite{schad2010runtime}. The study addresses that performance unpredictability in the cloud is a signiﬁcant issue for many users and is often considered a key obstacle in cloud adaptation. The study ﬁnds that Amazon EC2 shows high variance in its performance. The performance of clouds for scientiﬁc computing is analyzed using micro-benchmarks and kernels on Amazon EC2 in \cite{ostermann2009performance,iosup2011performance}. The proposed study observes that tested clouds are not suitable for scientiﬁc computing due to their performance variance and low reliability. A performance testing approach is proposed in \cite{he2021performance} to deal with unknown cloud performance variability.  

To the best of our knowledge, most existing studies do not consider IaaS performance variability for the long-term period. An extensive study on the variability of IaaS performance is carried out in \cite{leitner2016patterns}. The study suggests that cloud performance is a ``moving target" and requires re-evaluation periodically. A signature-based selection of IaaS cloud services is proposed in \cite{fattah2020icws}. The proposed work models the long-term performance variability of IaaS cloud services using the concept of signature. The signature of IaaS services is generated from the experience of the past trial users who share their data with a trusted third party. The trusted third party analyzes the periodic performance behavior of an IaaS service to generate its signature. However, the proposed work does not consider the changes in the signature over a long period of time \cite{fattah2020icws}. 

Change detection is an important research topic that identifies abrupt changes in a process \cite{veeravalli2014quickest}. It has been applied to many domains, including climate change detection, speech recognition, activity recognition, and edge detection in image processing. Existing approaches for the change detection problem are categorized as either ``offline" or ``online" methods \cite{aminikhanghahi2017survey}. Offline methods analyze the entire data set at once and find where the change had occurred. Online methods for change detection monitor and analyze each data point as they become available from a stream or source. Online methods typically rely on the statistical properties of the process to determine the change. We identify three criteria to evaluate change point techniques: a) ability to detect changes, (b) accurately identifying the change points, and (c) the number of tests to detect changes.

The \textit{experiences of existing users} are leveraged to estimate QoS performance of cloud services \cite{wang2019qos,yang2018location,tang2016collaborative}. The ranks of the providers are then measured based on the predicted performance. Collaborative filtering (CF) is a well-known approach to predict the QoS performance of a service based on the experience of existing users. A CF-based approach is proposed where observed QoS performance by past users is utilized to estimate QoS performance of a service for a new user \cite{wang2019qos}. During the prediction, the proposed approach utilizes similar users' experiences to estimate QoS performance. Similar users are selected based on a user's QoS requirements. A set of CF-based approaches is carried out where locations of users are utilized to calculate the similarity between the trial users \cite{yang2018location,tang2012location, shao2007personalized}. These studies show that the location of the user has a considerable effect on the QoS performance of a service. Most existing CF-based approaches primarily deal with short-term prediction and do not take into consideration the change in performance behavior in the long term.

To the best of our knowledge, there is no prior work that addresses the long-term IaaS performance change detection problem \cite{fattah2019long}. The proposed approach in \cite{fattah2020event} mainly focuses on the change point detection (CPD) in IaaS performance. The CPD is a pre-requisite of IaaS performance change detection \cite{aminikhanghahi2017survey}. In the CPD problem, the distribution of data before and after the change is often considered known. The proposed work in \cite{fattah2020event} introduces an ECA model to detect change points in IaaS performance behavior. The ECA approach is an effective CPD technique. Other change point detection techniques include Bayesian change point detection, Shapelet, Model fitting, and Gaussian process. The work in \cite{fattah2020event} utilizes the CUSUM control chart to detect changes in IaaS performance. CUSUM relies on the mean and standard deviation of a time series to detect changes. However, CUSUM is unable to differentiate between noise and change in IaaS performance \cite{page1961cumulative}. Change detection in time series data is usually performed using different similarity measure techniques. However, most of these approaches do not consider the noise that may appear in the data. Therefore, we introduce a change detection framework that identifies noise in IaaS performance by leveraging existing similarity measures and noise detection techniques.

\section{Conclusion}
Detecting changes in long-term IaaS performance is important as it will help new consumers to select the best services according to their long-term QoS requirements. We represent the long-term performance of an IaaS service using the concept of the IaaS performance signature. We propose a novel framework to detect changes in the performance behavior of an IaaS service as represented by its signature.  The key challenge in performance change detection is to differentiate between noise and changes in IaaS performance. We introduce a performance noise model where three types of noise are defined for IaaS performance behavior, i.e., spikes, attenuation, and distortion. We utilize time series similarity measure techniques and a sliding window technique to identify noise in IaaS performance. We proposed an SNR-based approach to improve the performance of change detection using prior knowledge about the noise. In this approach, we utilize the Signal to Noise Ratio to measure the level of noise in IaaS performance to detect changes. The experiment results show that the proposed framework detects changes in IaaS performance effectively. In future work, we aim to investigate IaaS performance noise in more detail to develop a more accurate change detection approach.

\section{Acknowledgement}
This research was partly made possible by LE220100078 and DP220101823 grants from the Australian Research Council. The statements made herein are solely the responsibility of the authors.

\bibliographystyle{ACM-Reference-Format}
\bibliography{Updated_Main}

\end{document}